\documentclass[usenatbib]{mn2e}
\usepackage{apjfonts,amsfonts,amsmath,amssymb,bm,ctable,verbatim}

\newcommand{\msun}{M_{\sun}}

\newcommand{\GIZMO}{{\small GIZMO}}
\newcommand{\gizmourl}{\href{http://www.tapir.caltech.edu/~phopkins/Site/GIZMO.html}{\url{http://www.tapir.caltech.edu/~phopkins/Site/GIZMO.html}}}

\newcommand{\acknowledgments}[1]{\begin{small}\section*{Acknowledgments}\end{small}{\noindent #1}\vspace{5pt}}

\title[MFM Schr{\"o}dinger Methods]{A Stable Finite-Volume Method for Scalar-Field Dark Matter}

\author[Hopkins]{
\parbox[t]{\textwidth}{Philip F.~Hopkins$^1$} \vspace*{4pt} \\
$^1$ TAPIR, Mailcode 350-17, California Institute of Technology, Pasadena, CA 91125, USA. E-mail:phopkins@caltech.edu 
}

\date{}
\begin{document}
\maketitle

\begin{abstract}
We describe and test a family of new numerical methods to solve the Schr{\"o}dinger equation in self-gravitating systems, e.g.\ Bose-Einstein condensates or ``fuzzy''/ultra-light scalar field dark matter. The methods are finite-volume Godunov schemes with stable, higher-order accurate gradient estimation, based on a generalization of recent mesh-free finite-mass Godunov methods. They couple easily to particle-based N-body gravity solvers (with or without other fluids, e.g.\ baryons), are numerically stable, and computationally efficient. Different sub-methods allow for manifest conservation of mass, momentum, and energy. We consider a variety of test problems and demonstrate that these can accurately recover solutions and remain stable even in noisy, poorly-resolved systems, with dramatically reduced noise compared to some other proposed implementations (though certain types of discontinuities remain challenging). This is non-trivial because the ``quantum pressure'' is neither isotropic nor positive-definite and depends on higher-order gradients of the density field. We implement and test the method in the code \GIZMO. 
\end{abstract}


\section{Introduction}

Recently there has been a surge of astrophysical interest in numerical methods to solve the Schr{\"o}dinger-Poisson equation for scalar fields with high occupation number, driven primarily by recent interest in axionic, ``Bose-Einstein condensate'' (BEC), ``ultra-light'' scalar-field or ``fuzzy'' dark matter models. In these models the dark matter (DM) is a boson with mass $\sim 10^{-22} - 10^{-6}\,$eV which forms a ``heavy'' high occupation-number ``condensate'' (so acts like a massive ``cold'' DM candidate on large scales), but exhibits coherent quantum effects on smaller scales. At scales $\sim 10^{-22} - 10^{-20}$\,eV, the de Broglie wavelength of the DM reaches parsec to kpc scales, leading potentially to interesting astrophysical consequences \citep[for reviews, see][]{suarez:2014.review,hui:fuzzy.dm.review}. Therefore it is particularly interesting to couple quantum dynamics to traditional N-body gravity methods used for cosmological simulations. 

Many numerical methods to date are variations of that in \citet{mocz:2015.schrodinger.eqn.solutions.sph}: they noted that one can write the relevant equations in the Madelung form \citep[e.g.][]{spiegel:1980.madelung.equations}, which amounts to coupling the ``usual'' collisionless Euler equations with self-gravity (solved by $N$-body methods) to an additional ``quantum pressure tensor'' (QPT) $\bm{\Pi}$ (i.e.\ force $\propto \nabla \cdot \bm{\Pi}$). However the QPT is neither isotropic nor positive-definite and depends on first and second-derivatives of the density field (so the force depends on third-derivatives); as such it is difficult to devise accurate and numerically stable schemes. For example, the original \citet{mocz:2015.schrodinger.eqn.solutions.sph} scheme and most derivatives are based on an extension of the smoothed-particle hydrodynamics (SPH) equation-of-motion with $\bm{\Pi}$ evaluated at particle locations using standard SPH gradient estimators, but these prove to be noisy and numerically unstable, which is especially problematic when one wishes to follow non-linear collapse of cosmological halos. Meanwhile spectral methods \citep[e.g.][]{schwabe:2016.spectral,mocz:spectral.becdm} do not necessarily conserve mass or couple easily to standard $N$-body codes. Still other methods have been developed as well, including finite-point methods \citep{goldberg:implicit.schroedinger,Visscher:explicit.schroedinger,edgington:thesis} which are accurate and conserve mass but require a uniform, time-independent rectilinear mesh (e.g.\ a fixed Cartesian grid) and uniform (global) timestepping \citep[although these have been successfully applied to DM simulations in e.g.][]{schive:2014.interference.scalar.field.dm,2014PhRvD..90b3517U,kopp:vlasov.fuzzy.dm,2019PhRvD..99f3509L}.

In this paper, we therefore introduce a Lagrangian finite-volume Godunov formulation using higher-order accurate matrix-based gradient estimators, which leads to a family of detailed methods. We consider a series of test problems to demonstrate the stability and accuracy of the methods, in the code \GIZMO.\footnote{The public version of \GIZMO\ is available at: \gizmourl} 

\section{Numerical Methods}

\subsection{Equations Solved}

The dynamics of a self-gravitating Bose-Einstein condensate/scalar field at high occupation number are well-described by the Schr{\"o}dinger-Poisson equation (SPE), $i\,\hbar\,\partial_{t} \Psi = [-(\hbar^{2}/2\,m)\nabla^{2} + m\,\Phi({\bf x})]\,\Psi$, with particle mass $m$, external potential $\Phi$ which we will take to be the gravitational potential obeying $\nabla^{2}\Phi = 4\pi\,G\,(m\,|\Psi|^{2} + \rho_{\rm other})$ (including both self-gravity and other contributions), and wavefunction $\Psi = |\Psi|\,\exp{(i\,\theta)} = \Psi_{\rm R} + i\,\Psi_{\rm I}$. Absorbing the constant $m$ into $\Psi$ as $\Psi\rightarrow\sqrt{m}\,\Psi$, and taking the real and imaginary parts of the SPE gives a pair of equations which can be written in conservative form:
\begin{align}
\label{eqn:SPE} \partial_{t} \Psi_{\rm R} + \nabla \cdot \left( -\nu\,\nabla \Psi_{\rm I} \right) &= \frac{\Psi_{\rm I}}{2\,\nu}\,\Phi\ , \\ 
\nonumber \partial_{t} \Psi_{\rm I}  + \nabla \cdot \left( \nu\,\nabla \Psi_{\rm R} \right) &= -\frac{\Psi_{\rm R}}{2\,\nu}\,\Phi\ ,
\end{align}
where $\nu \equiv \hbar/(2\,m)$ and the $\Phi$ terms here appear as a ``source.'' 

With our definitions above (and by obvious analogy in the Poisson equation), the classical mass density is $\rho \equiv |\Psi|^{2}$. It is straightforward to manipulate Eq.~\ref{eqn:SPE} to obtain the probability/mass conservation equation, in terms of the probability current ${\bf j}$:
\begin{align}
\label{eqn:current} \partial_{t}|\Psi|^{2} &+ \nabla\cdot {\bf j} = \partial_{t}\rho + \nabla\cdot {\bf j} = 0 \ , \\ 
\nonumber {\bf j} &\equiv i\,\nu\,\left(\Psi\,\nabla \Psi^{\ast} - \Psi^{\ast}\,\nabla \Psi \right) = 2\,\nu\,\left( \Psi_{\rm R}\,\nabla \Psi_{\rm I} - \Psi_{\rm I}\,\nabla \Psi_{\rm R} \right)\ .
\end{align}

The Madelung transformation identifies Eq.~\ref{eqn:current} with the classical continuity equation $\partial_{t}\rho = \partial_{t}|\Psi|^{2} = -\nabla\cdot(\rho\,{\bf u})$ for the ``velocity'' ${\bf u} \equiv {\bf j}/\rho = 2\nu\,\nabla \theta$ (the group velocity of a wavepacket). We can then entirely replace Eq.~\ref{eqn:SPE} with a pair of equations including the continuity and ``momentum density'' equation for ${\bf j}=\rho\,{\bf u}$: 
\begin{align}
\label{eqn:euler} \partial_{t}{\bf j} &+ \nabla \cdot \left[ {\bf j}\otimes{\bf u} + \bm{\Pi} \right ] = -\rho\,\nabla\Phi \ , \\ 
\nonumber \bm{\Pi}  &\equiv \nu^{2}\,\left[ \frac{(\nabla\rho)\otimes(\nabla\rho)}{\rho} - \nabla\otimes(\nabla\rho)\right] \ .
\end{align}
Eq.~\ref{eqn:current}-\ref{eqn:euler} are identical to the classical Euler equations for $\rho$ and ${\bf u}$, up to the form of the quantum pressure tensor (QPT)  $\bm{\Pi}$.

For a de Broglie wavelength much smaller than the horizon there is no effect of ``quantum'' dynamics on the global expansion of the Universe and the correction terms to Eqs.~\ref{eqn:SPE}-\ref{eqn:euler} in an expanding Universe are identical to those for a normal fluid \citep[see][]{hui:fuzzy.dm.review}. 

\subsection{Volume Decomposition: Finite-Volume Formulation}

We can solve any of Eqs.~\ref{eqn:SPE}-\ref{eqn:euler} by modifying the Lagrangian, mesh-free finite-volume Godunov (specifically the ``meshless finite volume/mass'' or MFV/MFM) method from \citet{hopkins:gizmo} -- developed for fluid dynamics -- to treat the appropriate fluxes. For example, the QPT in Eq.~\ref{eqn:euler} is analogous to other anisotropic pressure terms already well-studied \citep[e.g.\ in kinetic MHD;][]{hopkins:gizmo.diffusion}. This retains the advantages of Lagrangian methods and (optionally) equal element masses for N-body solvers, while allowing for higher-order solutions to the normal fluid equations in a conservative manner. 

Briefly, each traditional $N$-body DM ``super-particle'' $a$ now becomes a mesh-generating point at coordinate ${\bf x}_{a}$ used for volume decomposition: there is a well-defined volume $\Omega_{a}$ associated with each $a$, and finite-volume quantities (e.g.\ mass, momentum) carried by the particle represent the integral over all DM particles in the domain $\Omega_{a}$ (rather than the mass/velocity of a Monte-Carlo ``super-particle''). If we operator-split the gravity/external potential terms (solved in the usual fashion below), then the remaining part of Eqs.~\ref{eqn:SPE}-\ref{eqn:euler} can all be written in conservative form. In a frame moving with arbitrary velocity ${\bf v}$, the equation for some volumetric quantity ${\bf q}$ with flux ${\bf F}$ becomes ${\rm d}{\bf q}/{\rm d} t + \nabla \cdot({\bf F} - {\bf q}\otimes{\bf v}) = 0$. The corresponding equation for the finite-volume ${\bf Q} \equiv \int_{\Omega} {\bf q}\,d^{3}{\bf x}$ is obtained from the usual finite-volume transformation:
\begin{align}
\frac{{\rm d}\,{\bf Q}_{a}}{{\rm d}t} & 
= \int_{\Omega_{a}}\frac{{\rm d}{\bf q}}{{\rm d} t}\,{\rm d}^{3}{\bf x}  
= \int_{\Omega_{a}} \nabla \cdot \left({\bf q}\otimes{\bf v} - {\bf F} \right) \,{\rm d}^{3}{\bf x}  \\
\nonumber & = \int_{\partial \Omega_{a}}\left( {\bf q}\otimes{\bf v} - {\bf F}  \right) \cdot {\rm d}{\bf A} 
= \sum_{b}\,\left[ \left({\bf q}\otimes{\bf v}\right)^{\ast}_{ab} - {\bf F}^{\ast}_{ab} \right]\cdot {\bf A}_{ab}
\end{align}
where ${\bf A}_{ab}$ is the (oriented) face area between $\Omega_{a}$ and neighboring volume $\Omega_{b}$, and $\left({\bf q}\otimes{\bf v}\right)^{\ast}_{ab}$ and ${\bf F}^{\ast}_{ab}$ are the interface values of $({\bf q}\otimes{\bf v})$ and ${\bf F}$. The problem is now defined by an exchange of a finite-volume quantity (e.g.\ mass, momentum) between neighbors across a face, so is manifestly conservative of that quantity.


In MFV/MFM, the second-order accurate effective face is given by:
\begin{align}
\label{eqn:mfm.face.area.def} {\bf A}_{ab} &\equiv  \bar{n}_{a}^{-1}\,\bar{\bm{\xi}}_{ab}({\bf x}_{a}) + \bar{n}_{b}^{-1}\,\bar{\bm{\xi}}_{ba}({\bf x}_{b})\\ 
\nonumber \bar{\bm{\xi}}_{ab}({\bf x}_{a}) &\equiv {\bf T}_{a}^{-1} \cdot {\bf x}_{ba}\, W({\bf x}_{ba},\,h_{a}) \\
\nonumber {\bf T}_{a} &\equiv \sum_{c}\,({\bf x}_{ca} \otimes {\bf x}_{ca}) \,W({\bf x}_{ca},\,h_{a}) 
\end{align}
where $\bar{n}_{a} \equiv \sum_{b}\,W({\bf x}_{ba}\, , \, h_{a})$ is the kernel-density defined local mesh-generating point number density, ${\bf x}_{ba}\equiv {\bf x}_{b} - {\bf x}_{a}$, and $W$ is the kernel function with scale-length $h_{a}$ set adaptively to match the local kernel-averaged inter-element separation ($h_{a} \equiv \bar{n}_{a}^{-1/3}$).\footnote{In this paper we will use a cubic spline for $W$, with maximum radius of compact support $H_{a} = 2\,h_{a}$; other choices have small effects on our results.}  The volume of domain $a$ is $V_{a}=h_{a}^{3}$, so the density within the domain is $\rho_{a} \equiv M_{a} / V_{a}$ (and other primitive quantities ${\bf q}_{a} \equiv {\bf Q}_{a}/V_{a}$). This is a ``smoothed'' Voronoi decomposition.

\subsection{Mesh Motion}

Following \citet{hopkins:gizmo}, we can define the mesh motion ${\bf v}$ as desired. For arbitrary ${\bf v}$, the method is analogous to the ``MFV'' method in \citet{hopkins:gizmo}, and obviously results in ``mass fluxes'' between resolution elements (changes in $\int_{\Omega_{a}} |\Psi|^{2}\,d^{3}{\bf x}$). Below, we note experiments using ${\bf v}=\mathbf{0}$, or ${\bf v} \equiv {\bf j}/|\Psi|^{2}$ (evaluated at each location ${\bf x}_{a}$), or ${\bf v}$ following the same trajectories that N-body ``particles'' would have under gravity alone (e.g.\ moving them as test particles according to the gravitational field). 

If we adopt ${\bf v} = {\bf u}$, however, then we can also implement an ``MFM''-like method. Specifically, in the Reimann problem where we solve for ${\bf F}^{\ast}_{ab}$, we take the face velocity to be that of the contact wave $S_{\ast}$ between $a$ and $b$. As shown in \citet{hopkins:gizmo}, for ${\bf v}={\bf u}$ this is consistent to the second-order integration accuracy of the code, and leads (by definition) to exactly zero ``mass flux'' (volume-integrated flux of $\rho$)  between $a$ and $b$, so resolution elements retain exactly their original total mass ($\int_{\Omega_{a}} |\Psi|^{2}\,d^{3}\,{\bf x}$). This is especially convenient for coupling to N-body gravity solvers. However, because this exactly determines the solution of the mass flux/probability current (Eq.~\ref{eqn:current}), and depends on a well-defined ${\bf u}$, it can only be applied self-consistently if we solve the SPE in Madelung form (Eqs.~\ref{eqn:current}-\ref{eqn:euler}), as opposed to directly evolving $\Psi$ (Eq.~\ref{eqn:SPE}).

\subsection{Gravity \&\ Force-Softening}

As noted above, gravity is operator-split. This can be solved by the usual N-body methods, with one important improvement: since we have already de-composed the volume into a well-defined continuous density field, we must solve the gravity equations for the same density field. This amounts to using the fully-adaptive gravitational softening method (where the softening description is equivalent to the solution of the Poisson equation for the continuous density field) as described in detail in \citet{hopkins:gizmo,hopkins:fire2.methods} for the DM. If we are explicitly evolving ${\bf u}$ or $\rho\,{\bf u}$, this amounts to updating ${\bf u}$ according to $\nabla \Phi$ in the usual manner over a timestep $\Delta t$. If we explicitly evolve $\Psi$, we simply apply the unitary transformation $\Psi \rightarrow \Psi\,\exp{(-i\,\Phi\,\Delta t/2\,\nu)}$ over the same step. This already includes the corrections for the non-linear self-gravity terms as the domain described by a single element expands/contracts, so is manifestly conservative of momentum and energy \citep[see][]{price:2007.lagrangian.adaptive.softening}.

\subsection{Gradient Estimation: Calculating the Fluxes \&\ QPT}

Eqs.~\ref{eqn:SPE}-\ref{eqn:euler} depends on higher-derivatives of $\rho$, so it is critical to choose accurate, consistent, and stable gradient estimators. We therefore determine the first-derivative via the second-order accurate and consistent matrix gradient estimators
\begin{align}
\nabla_{a} f &= {\bf T}_{a}^{-1} \cdot \sum_{b}\,(f_{b} - f_{a})\,{\bf x}_{ba}\,W({\bf x}_{ba},\,h_{a})
\end{align}
This provides significant advantages in stability and accuracy over other numerical gradient estimation techniques \citep[for extensive discussion, see][]{maron:2012.phurbas.algorithm,luo:2008.compressible.flow.galerkin,lanson.vila:2008.meshfree.consistency,lanson.vila:2008.meshfree.convergence,hopkins:gizmo,mocz:2014.galerkin.arepo,pakmor.2016:improving.arepo.convergence}. For example, the gradient is robust to noise, as compared to face-centered gradient estimators, and its consistency does not depend on the details of the particle arrangement around the point $a$. As shown in \citet{gaburov:2011.meshless.dg.particle.method}, this is also the internally-consistent gradient operator for the MFM face operators.  We therefore use this to calculate $\nabla\rho$ and $\nabla\Psi$, then iteratively re-apply to each gradient component to calculate the ``gradients of gradients'' $\nabla\otimes\nabla\rho$ needed for $\bm{\Pi}$. This provides the required second-order accuracy, stability, and automatically expands the stencil by iteration  \citep[for examples and tests, see][]{munoz:2014.disk.planet.interaction.sims} -- in other words, since $\nabla_{a}\rho$ already involves a neighbor loop for each neighbor $b$, the neighbor stencil used for the second-derivatives is automatically (implicitly) larger, as needed to give robust results. Thus
$\bm{\Pi}_{a} \equiv \nu^{2}\,[\rho_{a}^{-1}\,(\nabla_{a}\rho)\otimes (\nabla_{a}\rho) - \nabla_{a}\otimes (\nabla_{a}\rho) ]$.


\subsection{Integration \&\ Timestepping}
\label{sec:timestep}

We integrate Eqs.~\ref{eqn:SPE}-\ref{eqn:euler} using the standard explicit, leapfrog scheme in \GIZMO\ using adaptive hierarchical timesteps (see \citealt{hopkins:gizmo} for details). The usual timestep critieria apply: e.g.\ acceleration-based criteria for N-body methods \citep[$\Delta t < \alpha\,(h_{a}/|{\bf a}_{a}|)^{1/2}$ with $\alpha\approx 0.4$;][]{power:2003.nfw.models.convergence}, and the Courant criterion needed for adaptive gravitational softening methods \citep[$\Delta t < 0.25\,{\rm min}((\nabla\cdot{\bf v})^{-1}_{a}\, , \, h_{a}/v^{\rm signal}_{a}$);][]{hopkins:fire2.methods}. But the SPE imposes its own timestep criteria. Regardless of how we parameterize it, because the SPE depends on higher numerical derivatives and the system admits whistler-type waves with $\omega = \nu\,k^{2}$, the fastest wavespeed is $\sim \hbar\,\Delta x/m \sim \hbar\,h_{a}/m$ and numerical stability under explicit integration require a timestep criterion of the form: 
\begin{align}
\label{eqn:timestep}\Delta t_{a} < C_{\rm CFL}\,\frac{m}{\hbar}\,h_{a}^{2}
\end{align}
where $C_{\rm CFL} = 0.25$ here. We emphasize this is critical for stability unless implicit integration is used, something which has not been widely recognized in this area. Quadratic timestep criteria (for e.g.\ whistler waves, diffusion/conduction problems, etc.) can be prohibitive at very high resolution (small $h_{a}$), motivating fully-implicit integration schemes, but for the DM problems of interest here the requirement is not particularly costly.

\subsection{The Riemann Problem \&\ Numerical Stability}
\label{sec:rp}

What remains is to define the interface flux (${\bf F}^{\ast}_{ab}$) to update ${\bf q}$. Following the MFV/M methods, this is given by solution of the appropriate Riemann problem at the face. We define left and right states with a piecewise-constant reconstruction, e.g.\ ${\bf F}_{R}={\bf F}_{a}$, ${\bf F}_{L}={\bf F}_{b}$ (while higher-order reconstructions are in principle straightforward, the already high-order derivatives in Eqs.~\ref{eqn:SPE}-\ref{eqn:euler} make this noisy and more difficult to stabilize). 
Following \citet{hopkins:gizmo}, it is both accurate and stable to approximate ${\bf F}^{\ast}$ using an HLL-type solver for an arbitrary tensor ${\bf F}$, well studied in elastic, fluid and collisionless (e.g.\ radiation) dynamics (this also allows for arbitrary equations-of-state and flux functions). We can write: 
\begin{align}
\label{eqn:reimann}{\bf F}^{\ast} &= \frac{S_{R}\,{{\bf F}_{L}} - S_{L}\,{{\bf F}_{R}}   + \alpha\,S_{R}\,S_{L}\,({\bf q}_{R} - {\bf q}_{L}) \cdot \hat{\bf A}}{ S_{R} - S_{L}  }  
\end{align}
where $S_{L,\,R}$ are initial wavespeeds and $\alpha$ is a limiter function. 

In the ``MFV-like'' methods \citep[see][]{gaburov:2011.meshless.dg.particle.method}, we take $({\bf q}\otimes{\bf v})^{\ast}_{ab} = {\bf v}_{\rm face}\otimes(S_{R}\,{\bf q}_{L}-S_{L}\,{\bf q}_{R})/(S_{R}-S_{L})$, with ${\bf v}_{\rm face} = ({\bf v}_{a}+{\bf v}_{b})/2$, and the wavespeeds are $S_{L} = {\rm min}(u_{\|}^{L} \, , \,u_{\|}^{R}) - c_{\rm eff}$, $S_{R} = {\rm max}(u_{\|}^{L} \, , \,u_{\|}^{R}) + c_{\rm eff}$ where $u_{\|}^{L,\,R} \equiv {\bf u}_{L,\,R} \cdot \hat{\bf A}$ and $c_{\rm eff}={\rm max}(c_{\rm eff,\,L},\,c_{\rm eff,\,R})$ is a charactistic wavespeed  \citep{toro:1999.reimann.solvers.book}. Because $c_{\rm eff}$ is the maximum wavespeed (below) so is always $\gg |u_{\|}|$, it makes a negligible difference if we simplify by adopting the Rusanov flux $S_{L,\,R}=\mp c_{\rm eff}$. 

Following \citet{hopkins:gizmo}, in ``MFM-like'' methods where ${\bf v}={\bf u}$ everywhere and we evolve Eqs.~\ref{eqn:current}-\ref{eqn:euler} in a finite-mass formulation, we take $v_{\rm face}\rightarrow S_{\ast}$, the contact wavespeed. Capturing this requires using the HLLC solver, where taking $v_{\rm face}\rightarrow S_{\ast}$ eliminates the advection equation and gives for the momentum equation (Eq.~\ref{eqn:euler}) $[{\bf F}_{ab}^{\ast} - ({\bf q}\otimes{\bf v})_{ab}^{\ast} ] \rightarrow \bm{\Pi}^{\ast}$ with the interface pressure:
\begin{align}
\label{eqn:reimann}\bm{\Pi}^{\ast} &= \frac{\tilde{w}_{R}\,\bm{\Pi_{L}} - \tilde{w}_{L}\,\bm{\Pi_{R}}   + \alpha\,\tilde{w}_{R}\,\tilde{w}_{L}\,(u_{\|}^{R} - u_{\|}^{L})\,\mathbb{I} }{ \tilde{w}_{R} - \tilde{w}_{L}  }  + \bm{\Pi}_{u}^{\ast}
\end{align}
with $\tilde{w}_{L,\,R} \equiv (S_{L,\,R} - u_{\|}^{L,\,R})\,\rho_{L,\,R}$.

The limiter $\alpha$ precedes the numerical diffusivity that arises from up-wind mixing in the Riemann problem. Some diffusivity is required for numerical stability, especially important here because $\bm{\Pi}$ can be negative and mesh-generating points move (which can source well-studied numerical instabilities such as the ``tensile instability'' from elasto-dynamics; \citealt{swegle:1995.sph.stability}). Taking $\alpha=1$ and $c_{\rm eff}$ equal to its maximum possible value (for a grid-scale particle separation $|{\bf x}_{ba}|$, $c_{\rm eff}^{\rm max} \sim (\hbar/m)\,(1/|{\bf x}_{ba}|)$) ensures unconditional stability, but produces a very diffusive scheme. This is especially un-desirable in the regime of interest when QPT forces are sub-dominant to gravity (one does not wish to artificially dissipate gravitational motions). By analogy to standard treatments of the tensile instability in negative-pressure tensors in elasto-dynamics \cite[e.g.][]{monaghan:2000.sph.tensile.instability} and previous studies of anisotropic viscous stress tensors in the MFM method \citep{hopkins:gizmo.diffusion}, we define
\begin{align}
\alpha &= \begin{cases}
	{\displaystyle 0} &  \hfill  (u^{L}_{\|} \le u^{R}_{\|})  \\
	{\displaystyle {\rm min}\left[ 1\, , \, \frac{\psi\,\| \bm{\Pi}^{\ast}_{\rm direct} \cdot {\bf A}\| }{\| \bm{\Pi}^{\ast}_{\rm diss} \cdot {\bf A} \|} \right]} &  \hfill  (u^{L}_{\|} > u^{R}_{\|}) 
\end{cases} 
\end{align}
where $\bm{\Pi}^{\ast}_{\rm direct} =(\tilde{w}_{R}\,\bm{\Pi_{L}} - \tilde{w}_{L}\,\bm{\Pi_{R}})/(\tilde{w}_{R} - \tilde{w}_{L}) $ and $\bm{\Pi}^{\ast}_{\rm diss} = (\tilde{w}_{R}\,\tilde{w}_{L}\,(u_{\|}^{R} - u_{\|}^{L})\,\mathbb{I} ) / (\tilde{w}_{R} - \tilde{w}_{L})$ represent the interface value without numerical dissipation and the dissipation term, respectively; 
$\| \bm{\Pi} \|^{2} \equiv \sum_{ij}  |{\Pi}^{ij}|^{2}$ denotes the Frobenius norm, and $\psi$ is a weight ($=10$ here, which is sufficient for stability without being too diffusive).\footnote{We estimate $c_{\rm eff} = (\hbar/m)\,k_{\rm eff}$ with $k_{\rm eff} = {\rm min}(|{\bf x}|_{ba}^{-1}\, , \, k_{\rm est}$), $k_{\rm est} = (1+\varpi)\,{\rm max}( |\nabla\rho|/\rho \, , \, |\nabla^{2}\rho| / |\nabla\rho| \, , \, \sqrt{|\nabla^{2}\rho_{a}-\nabla^{2}\rho_{b}|/(4\,|{\bf x}|_{ab}\,|\nabla\rho|)} )$ (evaluated with the interface gradient values). Here $\varpi_{ab} = \left[ (\bar{W}_{ab}/\bar{W}_{1/2})\,(\bar{H}_{ab}/|{\bf x}_{ab}|) \right]^{2}$, where $\bar{H}_{ab} \equiv (H_{a} + H_{b})/2$, $\bar{W}_{ab} \equiv (H_{a}^{3}\,W[{\bf x}_{ba},\,h_{a}] + H_{b}^{3}\,W[{\bf x}_{ba},\,h_{b}]) / 2$ is the dimensionless (pair-averaged) kernel function, and $\bar{W}_{1/2} \equiv H^{3}\,W(h\, , \, H)$ is the value of this function evaluated at the mean inter-element separation. }

The optional term $\bm{\Pi}_{u}^{\ast}$ is discussed in Appendix~\ref{sec:appendix.energy.conserving}, and exists to ensure energy conservation. We stress that the detailed form of $\alpha$ and $c_{\rm eff}$ have weak effects on our conclusions (entering only in the numerical dissipation, which becomes vanishingly small in converged solutions). But it respects three important properties: (1) it vanishes if the QPT does, (2) it vanishes for mesh-generating points which are well-separated/weakly interacting, or diverging/at rest (these only feel the QPT term), and (3) it rises rapidly (allowing the ``full'' numerical dissipation) for mesh-generating points which are approaching at small radii within the kernel.

\subsection{Different Method Variants}
\label{sec:flavors}

Thus far, we have defined our methods in generality without specifying {\em which} of Eqs.~\ref{eqn:SPE}-\ref{eqn:euler} we are actually solving. But of course, the SPE has only two degrees of freedom (e.g.\ $\Psi_{\rm R}$ and $\Psi_{\rm I}$) at a given $({\bf x},\,t)$, yet we have four evolution equations (one each for $\Psi_{\rm R}$, $\Psi_{\rm I}$, $\rho$ or $|\Psi |^{2}$, and ${\bf u}$ or ${\bf j}$). Analytically, the representations are equivalent. But numerically, this equivalence will be broken by truncation and integration errors. Thus we propose a family of closely-related methods, based on which quantities are explicitly evolved. Some methods within this family include:

\begin{enumerate}

\item{\bf ``Direct SPE'':} The simplest method {\em directly} evolves ${\bf q} = \Psi_{\rm R}$ and $\Psi_{\rm I}$, according to the finite-volume version of Eq.~\ref{eqn:SPE}. Thus we explicitly evolve the two components of the finite-volume quantity ${\bf Q} = \int_{\Omega} \Psi\,d^{3}{\bf x}$, and the fluxes for $\Psi_{\rm R,\,I}$ are just ${\bf F}_{\rm R,\,I}=\mp\,\nu\,\nabla \Psi_{\rm I,\,R}$. This will generally provide the most accurate evolution of complicated wave interference patterns through nodes. 
However, since the exchanged/conserved finite-volume quantities are the linear wavefunction components, while mass, energy, and momentum are {\em non-linear} combinations of these components and their gradients, this method does not automatically ensure manifest conservation or mass, momentum, or energy. These are conserved instead only up to {integration error}, because Eqs.~\ref{eqn:SPE}-\ref{eqn:euler} are equivalent analytically (i.e.\ at infinite resolution).\footnote{It is possible to construct unitary operators to evolve a discrete version of Eq.~\ref{eqn:SPE} such that mass is manifestly globally conserved \citep[see, e.g.][]{goldberg:implicit.schroedinger,Visscher:explicit.schroedinger}. However (to our knowledge) this has only been shown for very special configurations, e.g.\ finite-point methods with collocation on a uniform Cartesian mesh or one-dimensional moving mesh with a uniform (global) timestep, and the operators still do not maintain manifest momentum/energy conservation. Whether more general Lagrangian formulations allowing for irregular meshes or non-equal timesteps (common in cosmological simulations) can be formulated remains to be seen \citep[e.g.][]{twigger:thesis,edgington:thesis,Ceniceros:moving.mesh.schrodinger}.} 
In this method, we can set the mesh motion ${\bf v}$ to any desired values: in our experiments we have explored both ${\bf v}=\mathbf{0}$ and ${\bf v} = {\bf u} = {\bf j}/|\Psi|^{2}$.

\item{\bf ``Mass-Conserving SPE'':} In cosmological simulations, even small errors in DM mass conservation can be catastrophic (e.g.\ changing structure formation/cosmological expansion). A {\em manifestly} mass-conserving finite-volume formulation is obtained, however, if we explicitly evolve Eq.~\ref{eqn:current} for $|\Psi|^{2}$ or $\rho$ (i.e.\ resolution elements exactly exchange mass as a conserved quantity, with $m_{a} \equiv \int_{\Omega} |\Psi|^{2}\,d^{3}{\bf x}$), with flux ${\bf F}={\bf j}$ (written in terms of $\Psi_{\rm R}$, $\Psi_{\rm I}$). 
This gives three constraint equations for $\Psi$, so we cannot exactly update $\Psi_{\rm R}$ and $\Psi_{\rm I}$ independently as in method (i). In our implementation, we deal with this as follows. (1) Predict $\Psi_{\rm R}$ and $\Psi_{\rm I}$ (their finite-volume evolved quantities $\int_{\Omega}\,\Psi\,d^{3}{\bf x}$) from timestep 0 to 1 as $\Psi^{(0)} \rightarrow \Psi^{\prime}$ ($=\Psi_{\rm R}^{\prime} + i\,\Psi_{\rm I}^{\prime} = |\Psi^{\prime}|\,\exp{(i\,\theta^{\prime})}$) according to the ``Direct SPE'' method above (evolving Eq.~\ref{eqn:SPE}). (2) Use $\Psi^{\prime}$ to calculate the mass flux (${\bf F}={\bf j}$) and update the element mass $m^{(0)}\rightarrow m^{(1)}$ (hence $|\Psi|^{2} \equiv \rho^{(0)} \rightarrow \rho^{(1)}$) according to the manifestly mass-conserving finite-volume Eq.~\ref{eqn:current}. (3) Correct $\Psi^{\prime}\rightarrow \Psi^{(1)}$ conserving $|\Psi|^{(1)}$ (hence, mass) exactly, as $\Psi^{(1)} = |\Psi|^{(1)}\,\exp{(i\,\theta^{\prime})}$. Thus the {\em phase} of $\Psi$ is essentially evolved following method (i), while the {\em amplitude} of $\Psi$ is evolved directly as a finite-volume quantity, ensuring manifest mass conservation. 
Again, one can adopt any mesh motion ${\bf v}$, in principle. In our limited experiments, we have explored ${\bf v}=\mathbf{0}$, ${\bf v}={\bf u}$, and ${\bf v}=\bar{\bf u}$, where $\bar{\bf u}$ is a locally kernel-averaged value of ${\bf u}={\bf j}/\rho$ (smoothing ${\bf u}$ to avoid small-scale mesh motion and preventing divergences of ${\bf v}$ around nodes).

\item{\bf ``Momentum-Conserving / Madelung'':} Method (ii) promotes mass to a manifest, locally-conserved quantity. We can do the same for momentum, by explicitly evolving a conserved momentum ${\bf p}_{a} \equiv \int_{\Omega}{\bf j}\,d^{3}{\bf x}$ according to Eq.~\ref{eqn:euler}. Solving Eqs.~\ref{eqn:current} \&\ \ref{eqn:euler} for $m$ and ${\bf p}$ (or $\rho$ and ${\bf j}\equiv\rho\,{\bf u}$) means we are now solving the traditional Madelung form of the equations (there is no remaining degree-of-freedom for evolving $\Psi_{\rm R}$ or $\Psi_{\rm I}$, explicitly). 
This gives manifest mass and momentum conservation, but can introduce problems at ``nodes,'' where ${\bf u} \equiv {\bf j}/\rho$ diverges, as we discuss below. Because total energy is a non-linear combination of $\rho$, ${\bf u}$ (or $\Psi_{\rm R}$, $\Psi_{\rm I}$) and their gradients (independent of mass and momentum), however, this will only conserve energy to integration accuracy (see Appendix~\ref{sec:appendix.energy.conserving}). 
In this formulation, there is no ``penalty'' to making the method fully-Lagrangian (the difficulty with nodes is present, regardless), so we take ${\bf v}_{a}={\bf u}_{a}$ always, which allows us to adopt the finite-mass or ``MFM-like'' formulation which eliminates the mass equation entirely. Thus we only need to solve Eq.~\ref{eqn:euler} for the momentum (where here, we do not include the $\bm{\Pi}_{u}^{\ast}$ term).

\item{\bf ``Fully-Conservative'':} We can extend method (iii) further, and ensure manifest {energy} conservation as well, if desired. Naively this appears impossible, as there are only two degrees of freedom in $\Psi$ or ($\rho$, ${\bf u}$). We resolve this by introducing a new degree of freedom representing internal/un-resolved degrees of freedom of the micro-physical field (e.g.\ un-resolved oscillations in $\Psi$), analogous to the hydrodynamic internal energy or entropy. In our implementation here, this amounts to an additional field ($\bm{\Pi}_{u}^{\ast}$) which contributes in an isotropic manner to the QPT in Eq.~\ref{eqn:euler}, and is evolved according to its own finite-volume equation. Details and derivation of the added term are given in Appendix~\ref{sec:appendix.energy.conserving}. Again, we take ${\bf v}_{a}={\bf u}_{a}$, giving a ``fully Lagrangian'' or MFM-like formulation and eliminating the advection equation.

\end{enumerate}

We have experimented with all of the formulations (i)-(iv) above. For the rest of this paper, unless otherwise specified, we will restrict our tests to the ``Fully-Conservative'' formulation, as manifest conservation is perhaps the most unique and desireable feature of this formulation compared to other formulations of the SPE in the literature (especially in Lagrangian formulations). However, it is straightforward to adapt any of the methods into any other, and we will explore other variants in future work.

\begin{figure}
\includegraphics[width=0.95\columnwidth]{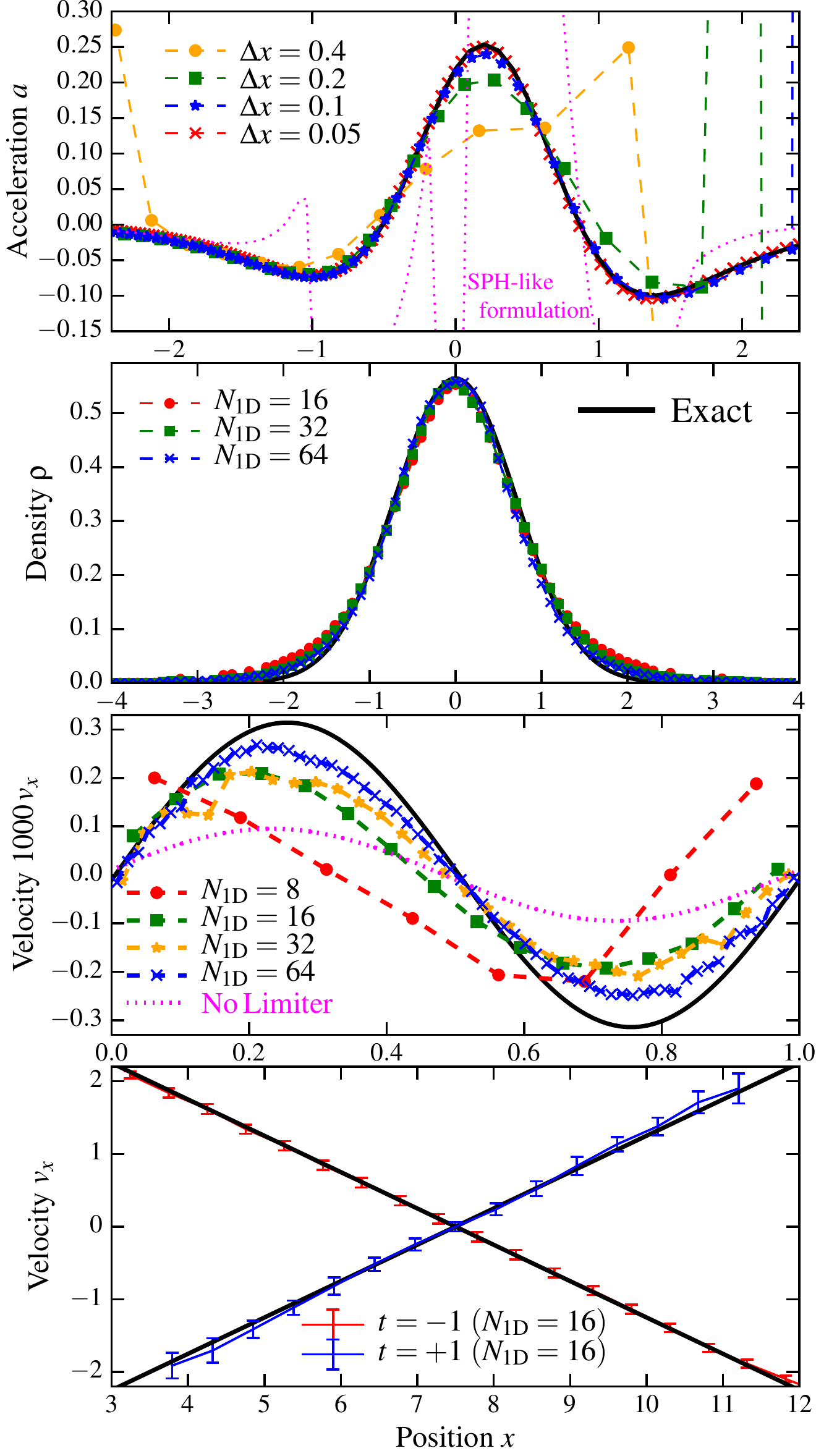} 
\vspace{-0.2cm}
\caption{
Test problems (\S\ref{sec:tests}; all run in 3D).
{\em Top:} Recovery of the exact acceleration (${\bf a} = -(\nabla\cdot\bm{\Pi})/\rho$) solution for a $\tanh$ density profile (\S\ref{sec:test:accuracy}; Eq.~\ref{eqn:tanh}). We compare exact analytic solution (black line) and the numerical acceleration computed for all particles by our scheme (points), as a function of resolution ($\Delta x$ is the mean inter-particle spacing in the $x$-direction over the plotted range). We also compare the popular ``SPH-like'' formulation of the QPT (Eq.~\ref{eqn:sph.like}), which is extremely noisy.
{\em Second:} Ground state of a damped harmonic oscillator after relaxation (\S\ref{sec:test:steady}). $N_{\rm 1D}$ is the equivalent one-dimensional particle number across the distance $x$ plotted.
{\em Third:} Traveling, oblique, linear whistler-type wave (\S\ref{sec:test:waves}), after two periods. ``No limiter'' shows an $N_{\rm 1D}=64$ run without a limiter to the numerical diffusivity in the Riemann problem (damping is much stronger). Runs with no numerical diffusivity, and all tested ``SPH-like'' formulations, at all are catastrophically unstable by this time.
{\em Bottom:} Converging-flow/intersecting-stream test (\S\ref{sec:test:nonlinear}) at initial ($t=-1$) and final ($t=1$) times, goes from converging (inflow) to diverging (outflow). Bars show $5-95\%$ range of all points.
}
\label{fig:tests}
\end{figure}

\section{Numerical Tests}
\label{sec:tests}

We now consider various test problems, all run using the identical code with the ``Fully-Conservative'' formulation described above. Unless otherwise specified, we do not include gravity or other forces and work in dimensionless units with $\hbar/m=1$ ($\nu=1/2$). All tests are run in 3D -- this is important for methods without a regular (e.g.\ rectilinear Cartesian) grid, where the error properties can be very different even in problems where the analytic solution depends on only one dimension, if particle positions vary in another dimension. We adopt equal particle masses, as (a) density variations require particle-position variations (e.g.\ an irregular mesh), which makes accurate gradient estimation more challenging, and (b) this is almost always the choice in cosmological simulations. 

\subsection{Instantaneous (Accuracy/Consistency) Tests}
\label{sec:test:accuracy}

First consider a non-dynamical test of the algorithm, to verify that it recovers (numerically) the correct accelerations. We initialize a density profile $\rho({\bf x})$ in a 3D box of arbitrarily large size, and save the numerically-calculated acceleration ${\bf a}_{a} = d{\bf u}_{a}/dt$ for each particle. 

{\bf Hyperbolic-Tangent Profile:} Following \citet{nori:2018.fuzzy.dm.gadget}, consider a profile varying along the $x$-axis with analytic form: 
\begin{align}
\label{eqn:tanh}\rho \propto 2 - \xi\ \rightarrow\ {\bf a} &= \frac{(1-\xi^{2})\,(7-\xi^{2}\{24+\xi [3\,\xi-16] \})}{4\,(2-\xi)^{3}}\,\hat{x}
\end{align}
where $\xi \equiv \tanh{(x)}$. So the density varies from some $\rho_{0}$ at $x \ll -1$ to $\rho_{0}/3$ at $x \gg +1$, over a width of $\sim 1$ in $x$. Fig.~\ref{fig:tests} plots the recovered discrete ${\bf a}_{a}$ at various resolutions. Even this simple density profile produces a complicated ${\bf a}$ which is not symmetric and has multiple features. With a few resolution elements ``across'' the density jump, our method recovers the main qualitative features; we see rapid convergence with decreasing $\Delta x$. 

For comparison, we show the results using the ``SPH-like'' formulation in \citet{mocz:2015.schrodinger.eqn.solutions.sph,velmaat:2016.cosmo.fuzzy.dm,zhang:2018.fuzzy.dm.implementation,nori:2018.fuzzy.dm.gadget}.\footnote{In \citet{nori:2018.fuzzy.dm.gadget}, 
\begin{align}
\nonumber \nabla_{\rm SPH} \rho_{a} &\equiv \sum (\rho_{b}-\rho_{a})\,\frac{m_{b}\,\nabla W_{ab}}{\sqrt{\rho_{a}\rho_{b}}} \\ 
\nonumber \nabla^{2}_{\rm SPH} \rho_{a} &\equiv - \frac{|\nabla_{\rm SPH} \rho_{a}|^{2}}{\rho_{a}} + \sum (\rho_{b}-\rho_{a})\,\frac{m_{b}\,\nabla^{2} W_{ab}}{\sqrt{\rho_{a}\rho_{b}}} \\ 
\label{eqn:sph.like} {\bf a}_{\rm SPH} &\equiv \frac{\hbar^{2}}{4\,m^{2}}\,\sum \left(\nabla^{2}_{\rm SPH} \rho_{b} - \frac{|\nabla_{\rm SPH} \rho_{b}|^{2}}{2\,\rho_{b}} \right)\,\frac{m_{b}\,\nabla W_{ab}}{f_{b}\,\rho_{b}^{2}}
\end{align}
where $\nabla W_{ab}$ refers to the gradient of the kernel function $W$ evaluated at ${\bf x}_{ba}$, and $f_{b} \equiv 1 + (H_{b}/3\rho_{b})\,\partial \rho_{b}/\partial H_{b}$ is the usual SPH correction for variable smoothing lengths \citep{springel:entropy}. As discussed in \citet{nori:2018.fuzzy.dm.gadget}, a variety of SPH formulations are possible (see e.g.\ \citealt{price:2012.sph.review,hopkins:lagrangian.pressure.sph} for general discussion of these degrees-of-freedom), but of those they consider this was the most accurate so we compare it here. We have specifically also tested the variant formulations in \citet{mocz:2015.schrodinger.eqn.solutions.sph,velmaat:2016.cosmo.fuzzy.dm,zhang:2018.fuzzy.dm.implementation} and find they all exhibit comparably poor gradient recovery and are all numerically unstable in dynamical tests.}
Even at high resolution, the SPH-like result is noise-dominated and systematically biased (as \citealt{nori:2018.fuzzy.dm.gadget} also showed). This owes to well-known problems: the  SPH gradient estimators and equation-of-motion (${\bf a} \propto \nabla\cdot \bm{\Pi}$) cannot be made zeroth-or-first order consistent/accurate without being unstable \citep{price:2012.sph.review}, and higher-order SPH gradients (especially the $\nabla^{2} W$ form) are notoriously noisy. 

{\bf 3D Gaussian:} We have also compared the 3D profile: 
\begin{align}
\rho \propto (1-\xi)^{-1} \rightarrow \ 
{\bf a} = \frac{\xi\,r}{4}\left(5-4\,\xi - r^{2}\,[1-\xi]^{2} \right)\,\hat{r}
\end{align}
where $\xi^{-1} \equiv 1 + \exp{(r^{2}/2)}$ and $r^{2}=|{\bf x}|^{2}$. Our results for this test are qualitatively identical to the $\tanh$ profile so we do not show both.

\begin{figure}
\begin{tabular}{r}
\vspace{-0.7cm}
\includegraphics[width=0.96\columnwidth]{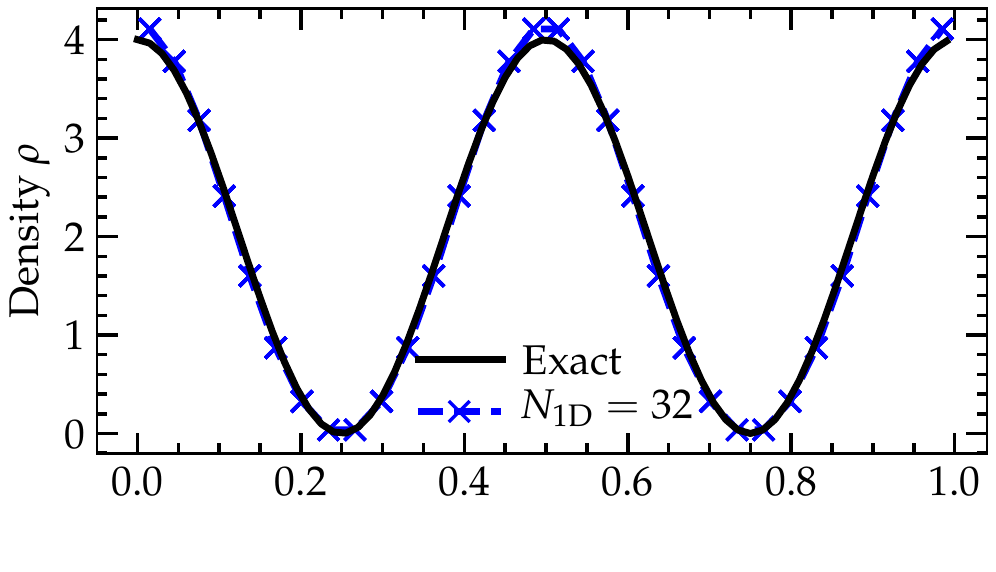} \\
\vspace{-0.7cm}
\includegraphics[width=0.96\columnwidth]{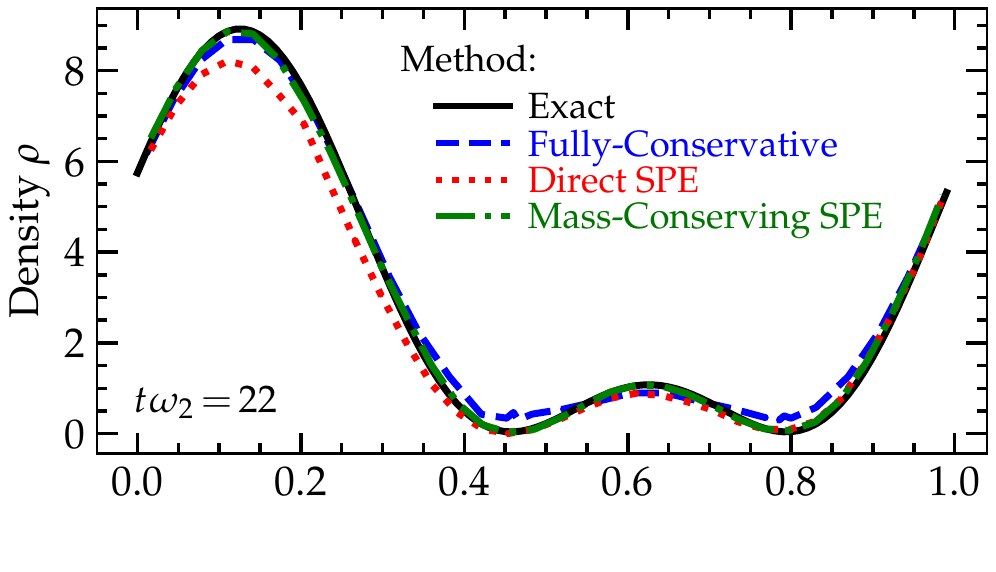} \\
\vspace{-0.7cm}
\includegraphics[width=0.96\columnwidth]{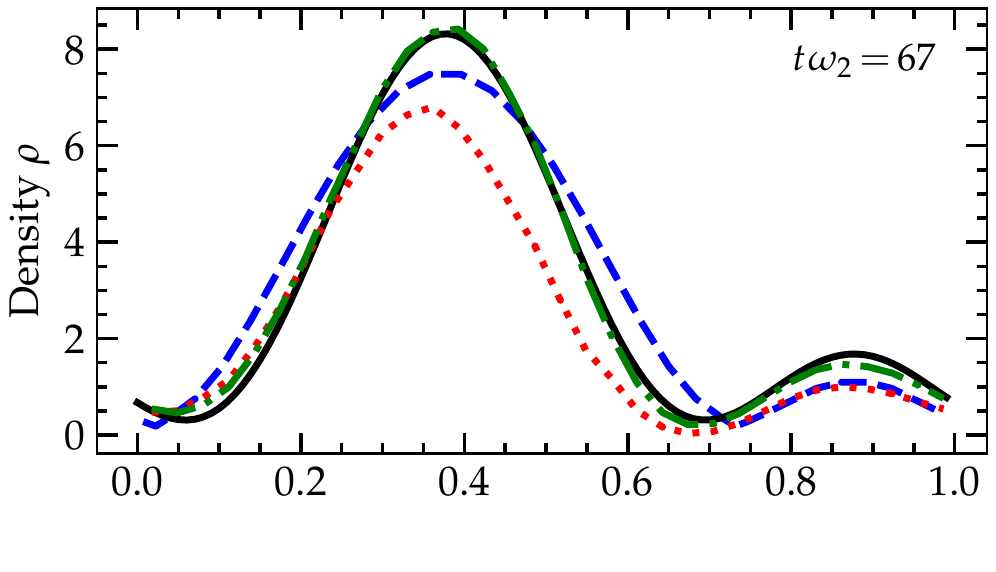} \\
\vspace{-0.0cm}
\includegraphics[width=0.96\columnwidth]{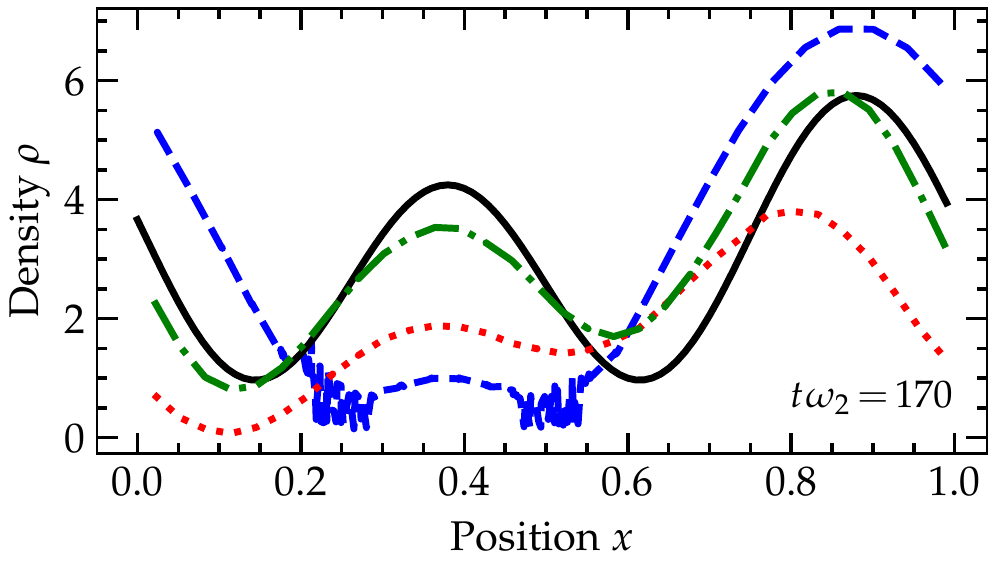} 
\end{tabular}
\vspace{-0.2cm}
\caption{
Wave interference-pattern tests with ``nodes'' where $\rho\rightarrow0$ (\S\ref{sec:nodes}). 
{\em Top:} A ``well-behaved'' node with $N_{w}=2$ interfering waves producing a traveling wave pattern with nodes, after $10$ periods (with $a_{2}=a_{1}=1$, $k_{2}=2\,k_{1}=4\pi$). The velocity ${\bf u}$ evolves continuously through the node, allowing our method to integrate it accurately for many periods, with convergence similar to the linear wave test. 
{\em Second-through-Bottom:} A ``dis-continuous'' node with $N_{w}=3$ ($a_{1,2,3}=1$, $k_{1,2,3}=(2\pi,\,4\pi,\,6\pi)$, $N_{\rm 1D}=32$), at various times from $t = (0-200)\,\omega_{2}^{-1}$ ($\sim 35$ periods of the fastest wave). At $t=0$ and every subsequent $\Delta t \sim 25\,\omega_{2}^{-1}$ interval, a pair of dis-continuous nodes (where ${\bf u}$ diverges and switches-signs discontinuously from $-\infty$ to $+\infty$) appear and disappear. After a few such intervals, the solution begins to diverge from exact (as dissipative terms dominate the evolution through the discontinuity). We compare other methods from \S~\ref{sec:flavors}: our ``Momentum-Conserving/Madelung'' formulation behaves similarly. ``Direct SPE'' evolves smoothly through nodes but numerical dissipation damps $|\Psi|^{2}$ (i.e.\ mass) monotonically. ``Mass-Conserving SPE'' provides a compromise which conserves mass and behaves well here.}
\label{fig:nodes}
\end{figure}

\subsection{Static/Steady-State Tests}
\label{sec:test:steady}

Now add full time-evolution, but consider steady-state solutions. 

{\bf Ground State of a Damped Harmonic Oscillator:} Particles are laid randomly (from a uniform PDF) in a periodic cube with side-length $L=8$ and mean density $\langle \rho \rangle = 1/L^{3}$, with uniform random velocities from $-10$ to $10$ in each direction. We add to the acceleration a term ${\bf a} \rightarrow {\bf a} - x\,\hat{x} - \gamma\,{\bf u}$ with $\gamma=4$, corresponding to a 1D harmonic oscillator potential ($V=x^{2}/2$) with frictional damping ($\gamma$) to force decay to the ground state. Fig.~\ref{fig:tests} plots the density distribution well after all velocities have decayed, and the exact ground state $\rho = (1/L^{2}{\pi^{1/2}})\,\exp{(-x^{2})}$. 

{\bf Advection of a (1D) Density Mode:} We initialize $\rho_{0}(x) = 1 + x + x^{2}/4$ in an arbitrarily large box, with uniform velocity ${\bf u}_{0}=(1,\,-1/\sqrt{3},\,1/\sqrt{2})$; this has ${\bf a}={\bf 0}$ so should advect as $\rho({\bf x},\,t) = \rho_{0}(x - t)$. Comparing at $t=0-10$ shows good agreement; because the $\rho$ profile is smooth (a second-order polynomial), any numerical inaccuracy in the initial ${\bf a}$ is small (much smaller than our $\tanh$ test), so ${\bf a}$ is vanishingly small (any initial noise from the finite-sampling of $\rho$ is resolved within a few timesteps by the diffusion in the Riemann problem). Since our Lagrangian method trivially solves advection and is manifestly Galilean-invariant, this is not a challenging test. 

\subsection{Linear/Wave Dynamics}
\label{sec:test:waves}

{\bf Oblique Traveling (Whistler-Type) Waves:} Consider Eq.~\ref{eqn:euler} with $\rho = \rho_{0}\,(1 + \delta\tilde{\rho})$, ${\bf u} = {\bf u}_{0} + \delta{\bf u}$ where $\rho_{0}$, ${\bf u}_{0}$ are homogeneous, and linearize in the perturbed quantities then make the usual Fourier {\em ansatz}: $\delta X({\bf x},\,t) = \delta X_{0}\,\exp{\{ i\,{\bf k}\cdot({\bf x} - {\bf u}_{0}\,t) - i\,\omega\,t \}}$. The linearized equations feature one non-trivial eigenmode: $\delta{\bf u} = (\omega/k)\,\hat{\bf k}\,\delta \tilde{\rho}$ with $\omega=\pm (\hbar/2 m)\,k^{2}$. So initialize an eigenmode: $\delta\tilde{\rho}_{0} = \epsilon\,\sin{(k\,x)}$ with $\hbar/m=1$, $\delta{\bf u}=(k/2)\,\hat{x}\,\delta\tilde{\rho}$ and $\epsilon=10^{-3}$, $k=2\pi$, in a 3D periodic box with side-length unity, ${\bf u}_{0} = (1,\,-1/\sqrt{3},\,1/\sqrt{2})$, and the particles laid on an initially Cartesian grid rotated by the arbitrarily-chosen Euler angles $(41^\circ , \, -23^\circ , \, 67^\circ )$ before being perturbed to generate the density profile, so that both ${\bf k}$ and ${\bf u}_{0}$ are oblique to the ``grid'' and each other. We evolve the system for 40 wave periods ($\omega\,t = k^{2}\,t/2 = 20\pi$).

Our implementation produces convergence and numerical stability. We have repeated this with $k=4\pi,\,8\pi$ and different oblique angles or ${\bf u}_{0}$ and obtain essentially identical results. As usual, the numerical dissipation in the Riemann problem produces some damping, but this decreases at higher resolution (as desired). Without dissipation ($\alpha=0$ in Eq.~\ref{eqn:reimann}), the low-resolution solutions are closer to exact at early times (since they are undamped) but we see numerical instability set in later and destroy the solution. We have also verified that the solutions become numerically unstable if we remove the quadratic timestep condition (\S~\ref{sec:timestep}), as expected. With the HLLC diffusion term but no limiter ($\alpha=1$, $k_{\rm eff}=1/|{\bf x}_{ab}|$), the wave is always damped, even at high resolution (because the grid-scale wavespeed $c_{\rm eff}$ increases at high resolution as $\sim 1/\Delta x$, one cannot rely on resolution alone without a limiter here).

{\bf Simple Harmonic Oscillator:} We initialize $\rho_{0}(x) = (1/L^{2}{\pi^{1/2}})\,\exp{(-x^{2})}$, ${\bf u}_{0} = (1,\,-1/\sqrt{3},\,1/\sqrt{2})$, and add to the acceleration equation the term ${\bf a} \rightarrow {\bf a} - x\,\hat{x}$ (corresponding to $V=x^{2}/2$). This has exact solution $\rho({\bf x},\,t) = \rho_{0}(x - \sin{[t]})$. Although fully non-linear, we find this is a much ``easier'' test than the whistler-type waves above, as (1) the dispersion relation is simpler (acoustic-like, rather than whistler-like, making it much easier to stabilize), (2) there is less ``structure'' in the wave, (3) the external potential ``corrects'' the system (as opposed to being self-generating), and (4) the large (non-linear) amplitudes of the gradients make it somewhat less sensitive to numerical noise. Therefore we do not show the results but simply note the convergence is substantially faster than for the whistler-type wave.

\subsection{Strongly Non-Linear Tests}
\label{sec:test:nonlinear}

{\bf Two-Stream Intersection:} Following \citet{hui:fuzzy.dm.review} consider the non-linear evolution of a converging flow with a Gaussian super-position of plane waves as a convenient (and exactly analytically solveable) proxy for the intersection of oppositely-moving streams. In a large periodic, cubic box, $\rho = 1/\sqrt{\pi\,f}\,\exp{(-x^{2}/f)}$ and ${\bf u} = (-t\,x/f\, , \, 0 \, , \, 0)$ with $f=1+t^{2}$ is an exact solution at all times $t$. We initialize this at $t=-1$, corresponding to a converging flow on the origin, and run it until $t=+1$, at which point it should be exactly mirrored as a diverging flow (representing the plane waves collisionlessly ``passing through'' one another, or equivalently a ``bounce'' as the quantum pressure diverges when the converging flow collapses). The scheme is able to recover the diverging solution, which demonstrates it can capture intersecting streams and quasi-collisionless phenomena, and also that the numerical dissipation is not so large to damp the motion (preventing the ``bounce'' and simply merging into a non-moving soliton at the origin). 

{\bf Conservation Test:} To ``stress-test'' conservation and stability in the code, we initialize a periodic box with side-length, mass, gravitational constant, and $\hbar/m$ equal to unity, laying particles down randomly according to a Poisson distribution with random initial velocities drawn from a 3D Gaussian with dispersion $=10^{3}$ (with zero net momentum). We evolve this, including self-gravity, until time $=100$. We verify (1) the code remains numerically stable, (2) timesteps and densities remain positive-definite, (3) mass is (trivially) conserved, (4) linear momentum is manifestly conserved to machine accuracy if we adopt a single timestep for all particles, (5) with adaptive (individual) timesteps, variations in timestep between neighbors mean the simple implementation of our scheme is not machine-accurate momentum-conserving, but we find the linear momentum error $|\sum {\bf p}_{a}| / \sum |{\bf p}_{a}|$ is less than $1\%$ at all times.

\begin{figure}
\begin{tabular}{r}
\includegraphics[width=0.86\columnwidth]{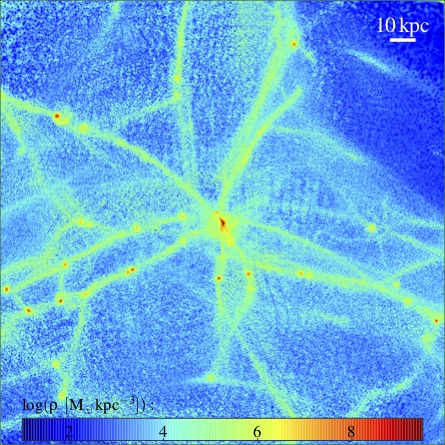} \\
\includegraphics[width=0.96\columnwidth]{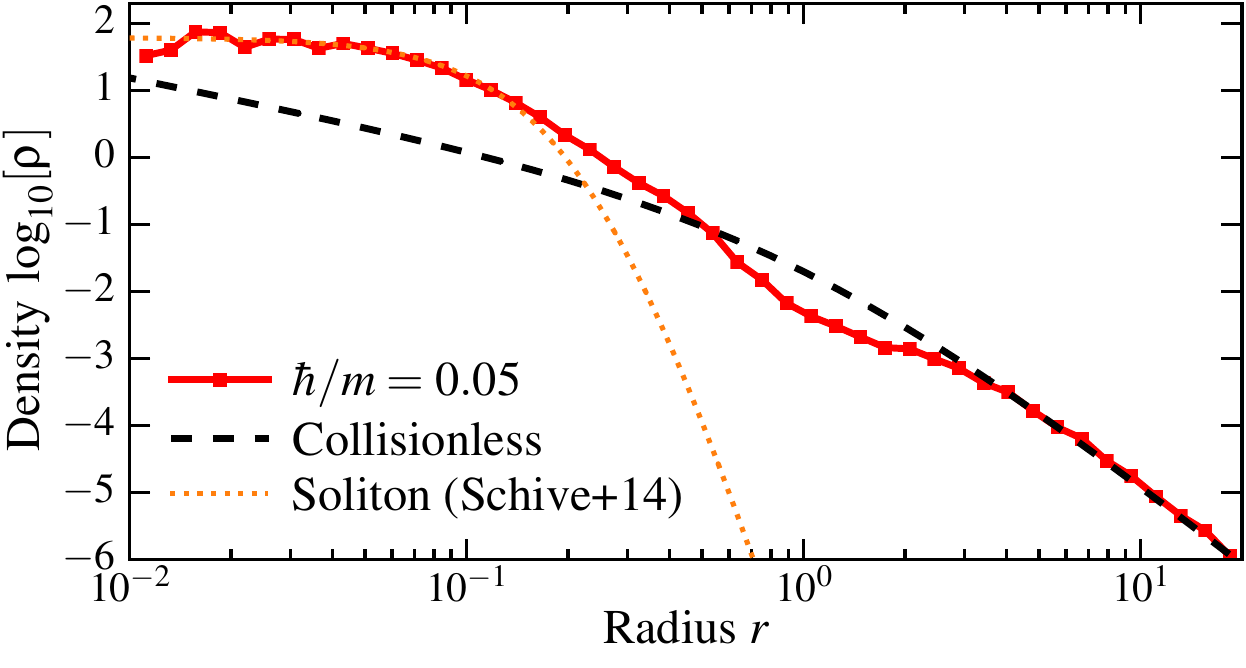} 
\end{tabular}
\vspace{-0.2cm}
\caption{
Dark matter halo tests (\S\ref{sec:test:dm}). 
{\em Top:} DM density map at $z=2.5$ in fully-cosmological, high-resolution ``zoom-in'' simulations (from $z=100$) of a $10^{10}\,\msun$ region, with $m\,c^{2} = 10^{-21}\,$eV. ``Graininess'' (substructure from wave interference) and solitons at halo centers are evident; the method is able to handle non-linear cosmological integration stably.
{\em Bottom:} Self-gravitating DM halo (density profile shown). The halo starts from an equilibrium phase-space distribution for {\em collisionless} particles (labeled), but then the QPT is ``turned on'' and the system relaxes to a new ``ground state'' with a ``soliton'' in the center of size $\sim \hbar/m \sim 0.05$. Dotted line shows the best-fit soliton profile following \citet{schive:2014.interference.scalar.field.dm}.
}
\label{fig:dm}
\end{figure}

\subsection{Interference Patterns \&\ the ``Node Problem''}
\label{sec:nodes}

In Fig.~\ref{fig:nodes}, consider an arbitrary $\Psi$ composed of a sum of $N_{w}$ independent one-dimensional plane waves $\Psi = \sum_{m}\,\psi_{m}$ where $\psi_{m} \equiv a_{m}\,\exp{\left[ i\,(k_{m}\,x - \omega_{m}\,t) \right]}$, $\omega_{m} = \nu\,k_{m}^{2}$. $N_{w}=1$ ``components'' gives $\rho=|a_{1}|^{2}$, $u=2\,\nu\,k_{1}$, i.e.\ uniform advection, which is trivially solved in our method. 

{\bf Well-Behaved Nodes:} With $N_{w}=2$, we obtain a traveling wave $\rho = \sum_{m}|a_{m}|^{2} + 2\,g_{12}$ where $g_{mn} \equiv a_{m}\,a_{n}\,\cos{[(k_{n}-k_{m})\,x - \nu\,(k_{n}^{2} - k_{m}^{2})\,t)]}$, and $u=j/\rho$ with $j=2\nu\,[\sum_{m}|a_{m}|^{2}\,k_{m} + (k_{1}+k_{2})\,g_{12}]$. This has non-trivial wave structure, but is just the non-linear version of the wave test from \S~\ref{sec:test:waves}. For $|a_{1}|/|a_{2}| \gg 1$ or $|a_{2}|/|a_{1}| \ll 1$ it reduces to a linear wave, but even if we choose $|a_{2}|=|a_{1}|$, so that the wave has ``nodes'' where $\rho = 0$ exactly (and the wave amplitude is fractionally as large as possible), this gives solutions comparable to the linear wave problem above, in their accuracy and rate of convergence.\footnote{For the $N_{w}=2$ problem, with $|a_{2}|=|a_{1}|$ so nodes exist, $u=\nu\,(k_{1}+k_{2})$ is constant even at nodes.} 

In other words, the mere existence of interference and nodes with $\rho=0$ does not necessarily cause problems for our method. Many other Madelung formulations fail catastrophically at nodes \citep[see][]{schive:2014.interference.scalar.field.dm,schwabe:2016.spectral,kopp:vlasov.fuzzy.dm,2019PhRvD..99f3509L}: the formulation here is more robust because (1) it evolves volume-integrated conserved quantities (e.g.\ momentum, mass) which remain finite; (2) the mesh moves with mass so can never evaluate Eq.~\ref{eqn:euler} exactly at a node but always outside (bracketing) said node; (3) numerical diffusion terms damp local divergences and prevent numerical instability at nodes; (4) strict conservation allows solutions to ``recover'' correct bulk properties even after integration through nodes.

{\bf Discontinuously-Divergent Nodes:} At $N_{w}=3$, $\rho=\sum_{m}|a_{m}|^{2} + 2\,\sum_{n>m}\,g_{mn}$, $j/(2\,\nu)=\sum_{m}|a_{m}|^{2}\,k_{m} + \sum_{n>m}(k_{m}+k_{n})\,g_{mn}$ so we have three-wave interference and velocity structure. If one of the $|a_{m}|$ is much larger or smaller than the others, then this reverts to the $N_{w}=1$ or $N_{w}=2$ case above with a superposition of linear waves, which add independently in our method, so the behavior is again comparable to the linear wave test above (the method is well-behaved). But if $N_{w} \ge 3$ with $|a_{1}|=|a_{2}|=|a_{3}|$ and $k_{1}\ne k_{2}\ne k_{3}\ne 0$, the interference pattern is challenging to capture. This is because a new type of ``node'' manifests where both $\rho \rightarrow 0$ and $u \rightarrow \pm \infty$ appear and disappear {\em dis-continuously} in time. Take $a_{1}=a_{2}=a_{3}$ and $(k_{2},\,k_{3})=(2,\,3)\,k_{1} \ne 0$ as a representative example. At almost any random time, $u$ is complicated but well-behaved (smooth, finite, and differentiable) {\em even} when there are nodes (these are ``well-behaved'' nodes, as above): in fact at $t=0$, $u=2\,\nu\,k_{1}$ is a constant, despite $\rho=0$ at $x = (2\pi/k_{1})\,(j \pm 1/3)$. But at {infinitesimally} smaller/larger $t = \pm \epsilon$ (at the same $x$), $u$ diverges as $u \propto \pm 1/\epsilon$, which then becomes well-behaved at later times. So going ``into'' certain times, $u$ diverges rapidly to $\pm\infty$ at certain points, then discontinuously ``jumps'' to a constant and then $\mp \infty$ (switching sign), before relaxing back ``out of'' the node, until the cycle repeats. Note, going to $N_{w}\gg 3$ we see no other qualitatively new ``node types'' appear. 

Our method can stably integrate ``through'' such a node, whereas the ``SPH-like'' methods, or our method with numerical dissipation terms turned off ($\alpha=0$ in \S~\ref{sec:rp}) immediately fail catastrophically. However, each integration ``through'' nodes degrades the solution, until after $\sim5-7$ such integrations qualitative features (e.g.\ peaks in $\rho$) are lost. Note that energy, mass, and momentum are still manifestly conserved, but higher-$k$ features in the interference pattern are lost. Of course, all numerical methods with finite integration error will diverge from an exact solution over time: what is especially problematic here is that convergence may be impossible. The issue is that a sharp discontinuity (in time and space) appears in the primitive variable ${\bf u}$ (which automatically reduces the theoretically ``best-case'' convergence rate for Godunov-type methods to first-order), and the magnitude of the discontinuity (in the physical solution) {\em increases} as $\sim 1/\Delta t \propto 1/\Delta x$ as the resolution ($\Delta t$ or $\Delta x$) improves. Because the discontinuity increases in size equal to or faster than the optimal convergence rate, the solution ``through'' the discontinuity will {\em always} (at any resolution) be dominated by the numerical dissipation terms. This is what drives the departure from the exact solution, until eventually a completely different solution is reached.\footnote{A less important, but still notable challenge with dis-continuous nodes is that while the total momentum is evolved, the mesh-generating points are advected with ${\bf u}$, so the signal velocity becomes very large and timesteps very small moving in/out of such nodes.} 

In principle, it may be possible to derive a different form of the numerical dissipation which preserves the ``correct'' analytic behavior through these dis-continuous nodes (essentially a better, problem-specific Reimann problem solution). But this is also where other methods in \S~\ref{sec:flavors} have advantages. Directly integrating $\Psi$ with our ``Direct SPE'' formulation, these dis-continuous nodes pose no challenge, because the wavefunctions $\psi_{n}$ are perfectly smooth and well-behaved at the nodes \citep[see][]{schive:2014.interference.scalar.field.dm,schwabe:2016.spectral,kopp:vlasov.fuzzy.dm,2019PhRvD..99f3509L}. However, as implemented here, the finite-volume method for $\psi$ is dissipative, so amplitudes damp -- meaning mass is monotonically dissipated. This converges with resolution (the damping is reduced), but is still not ideal -- better, unitary methods for integrating $\Psi$ would be preferred (see \S~\ref{sec:flavors}). Our ``Mass-Conserving SPE'' method presents one possible compromise, which behaves better on this particular problem and exhibits actual convergence (but sacrifices manifest momentum and energy conservation). The alternative ``Momentum-Conserving/Madeling'' formulation does not improve the behavior here, as it still evolves Eq.~\ref{eqn:euler}.

\subsection{Dark Matter Halos \&\ Cosmological Simulations}
\label{sec:test:dm}

{\bf Isolated Halo:} We initialize a spherically-symmetric, \citet{hernquist:profile} mass profile ($\rho = 1/[2\pi\,r\,(r+1)^{3}]$), self-gravitating ($G=1$) sphere with an isotropic velocity distribution function, using $10^{4}$ particles. For truly collisionless particles (no QPT), this is an exact equilibrium state. We then enable the QPT, with $\hbar/m = 0.05$. The potential at $r\rightarrow0$ is $-1$ in these units in the collisionless case (so velocity dispersion $v_{0} \sim 1$), and so the ground state of the halo with the QPT, while it has no closed-form solution, should approximately be a similar mass profile at large radii, with a constant-density soliton at the center of size $\sim \hbar/(m\,v_{0}) \sim 0.05$. We consider cases with and without an additional damping ${\bf a} \rightarrow {\bf a} - 4\,{\bf u}$: without damping the central density oscillates owing to its initial highly out-of-equilibrium central density, with damping the system quickly converges to a stable equilibrium ``ground state'' with these properties. Fig.~\ref{fig:dm} also compares the central mass profile to the specific ``soliton profile'' proposed in \citet{schive:2014.interference.scalar.field.dm}, $\rho \propto [1 + 0.091\,(r/r_{c})^{2}]^{-8}$, for which we find a best-fit $r_{c}\approx0.065$, similar to the order-of-magnitude expectation $\sim 0.05$ (with a slower fall at large $r$, as expected, as the ``collisonless-like'' or ``non-soliton'' component dominates; compare e.g.\ \citealt{mocz:spectral.becdm}).

{\bf Cosmological Simulation:} We consider a cosmological, DM-only ``zoom-in'' simulation of a $\sim 10^{10}\,\msun$ DM halo (at $z=0$)  in a $100\,$Mpc cubic volume, initialized at $z=100$ with a standard $\Lambda$CDM cosmology, and zoom-in (high-resolution) Lagrangian region including all particles within $\sim5$ virial radii of the halo at $z=0$, with particle mass $\sim 10^{4}\,\msun$ ($\sim10^{6}$ particles in the halo). The specific halo is {\bf m10q} with all details in \citet{hopkins:fire2.methods}. We use physical units with $m\,c^{2} = 10^{-21}$\,eV, which should produce a $\sim 0.1$\,kpc soliton in the halo center. We run to $z=2.5$ to verify that our method allows us to stably evolve scalar-field DM through non-linear cosmological history. A formal study of the solitons and halo mass density profiles that form in our cosmological simulations as a function of $\hbar/m$ and resolution (comparing some of the formulations above to literature results in e.g.\ \citealt{schive:2014.interference.scalar.field.dm,PhysRevD.98.043509,bar:fdm.rotation.curves.obs}) will be the subject of future work (Robles et al., in preparation).

\section{Discussion}

We have developed and tested a family of methods to solve the Schr{\"o}dinger-Poisson equation, relevant at high occupation number for simulations of e.g.\ Bose-Einstein condensates or axionic/ultra-light scalar-field (``fuzzy'') DM. The methods are based on the finite-volume, meshless hydrodynamic schemes in \citet{hopkins:gizmo}. They couple in a straightforward manner to particle-based N-body gravity solvers, are numerically stable, can trivially be solved with additional fluids present, and add only a fixed computational overhead for simulations which already compute gravitational forces with adaptive gravitational softenings. The particular variant tested here maintains equal particle masses, and manifestly conserves mass, momentum, and energy, though each of these is ``optional'': machine-accurate conservation can be traded for integration-error level conservation and more accurate treatment of e.g.\ rapidly-oscillating fields. We implement the methods in the code \GIZMO. 

Unlike some previous ``SPH-like'' implementations discussed in the literature, we show that the methods here recover gradients accurately and retain stability even given highly dis-ordered particle configurations. These are critical in chaotic situations including N-body gravity and cosmological simulations, but even in simple test problems (e.g.\ a traveling oblique wave) they are challenging, because the ``quantum pressure'' force ($\nabla \cdot \bm{\Pi}$) depends on first, second, and third derivatives of the density field, and the ``quantum pressure tensor'' ($\bm{\Pi}$) is neither isotropic nor positive definite. We present a series of test problems and show it is viable for high-resolution cosmological simulations of scalar-field DM. In future work, we intend to investigate these DM models in greater detail, in cosmological hydrodynamic simulations.

It remains to be seen how these methods (including the various ``flavors'' or sub-methods in \S~\ref{sec:flavors}, as opposed to just the ``Fully-Conservative'' formulation we have explored in more detail) compare with spectral and grid-based methods for solving the SPE, in various specific contexts (e.g.\ cosmological simulations). The most obvious advantage of the methods here in this context is their ability to be modularly built on top of standard N-body solvers, and their robust conservation properties. But they also have disadvantages of introducing numerical dissipation/diffusion, and noise from irregular, moving mesh configurations. Future work in these areas will shed light on the suitability of different numerical approaches for specific astrophysical problems of interest.

\acknowledgments{We thank Victor Robles, Michael Kopp, Xinyu Li, and our anonymous referee for a number of helpful discussions and suggestions. Support for PFH was provided by an Alfred P. Sloan Research Fellowship, NSF Collaborative Research Grant \#1715847 and CAREER grant \#1455342. Numerical calculations were run on the Caltech compute cluster ``Wheeler,'' allocations from XSEDE TG-AST130039 and PRAC NSF.1713353 supported by the NSF, and NASA HEC SMD-16-7592.}

\bibliography{ms_extracted}

\begin{appendix}

\section{Ensuring Energy Conservation With Un-Resolved Density Structure (Un-Resolved Quantum Potentials)}
\label{sec:appendix.energy.conserving}

A general problem arises when we wish to solve Eq.~\ref{eqn:euler} at finite resolution. The Schr{\"o}dinger-Poisson equation conserves energy, and this is manifest in the Madelung form (Eq.~\ref{eqn:euler}) if we define the total energy 
$E = \int d^{3}{\bf x}\,\rho\,[ {\bf u}\cdot {\bf u}/2 + \Phi + Q ]$, i.e.\ the sum of the usual kinetic and gravitational potential terms, plus the ``quantum potential'' $Q \equiv 2\,\nu^{2} (\nabla^{2}\sqrt{\rho})/\sqrt{\rho}$. But since $Q$ depends only on $\rho$, which depends only on the mass configuration, i.e.\ on the initial conditions and subsequent ${\bf u}({\bf x},\,t)$, then if there is any numerical error in integration, or non-vanishing numerical dissipation terms $\bm{\Pi}^{\ast}_{\rm diss}$, these cannot conserve energy (they change the velocities but not $\rho$ or, therefore, $Q$). In usual hydrodynamics, this is not a problem, because there is an additional thermodynamic variable (e.g.\ temperature, entropy), so although the numerical terms like $\bm{\Pi}^{\ast}_{\rm diss}$ may change the kinetic energy, this can always be exactly offset by an appropriate exchange with thermal energy (so the terms are {\em diffusive}, but {\em manifestly conservative}). 

In order to restore manifest energy conservation, we must introduce a ``thermodynamic-like'' variable into which the energy lost to (inescapable) numerical dissipation can be ``stored.'' This is the origin of the $\bm{\Pi}_{u}^{\ast}$ term in Eq.~\ref{eqn:reimann} -- it is (numerically) analogous to a standard thermodynamic pressure. Each element carries a scalar $\Pi^{u}$ (initialized to $0$), which evolves and defines $\bm{\Pi}_{u}^{\ast}$ according to: 
\begin{align}
\frac{d(V_{a}\,\Pi^{u}_{a})}{dt} &= \frac{1}{2}\,\sum_{b} ({\bf v}_{a} - {\bf v}_{b}) \cdot [ \bm{\Pi}^{\ast}_{\rm diss} + \bm{\Pi}_{u}^{\ast}] \cdot {\bf A}_{ab} \\ 
\bm{\Pi}_{u}^{\ast} &\equiv (\gamma_{\Pi}-1)\,\frac{(\tilde{w}_{R}\,\Pi^{u}_{L} - \tilde{w}_{L}\,\Pi^{u}_{R})}{(\tilde{w}_{R} - \tilde{w}_{L})}\,\mathbb{I}
\end{align}
Here $ \bm{\Pi}^{\ast}_{\rm diss} $ is a ``source term'' -- any work done by the numerical diffusion term is added to the ``energy'' $V_{a}\,\Pi_{a}^{u}$ (because of the form of $\alpha$, it is straightforward to show this is positive-definite). As $\Pi^{u}$ builds up, this generates an isotropic ``pressure'' $\bm{\Pi}_{u}^{\ast}$, which appears in the equation of motion (Eq.~\ref{eqn:reimann}), and therefore does $P\,dV$ work which must be added/removed from the energy. The coefficient $\gamma_{\Pi} = 5/3$, as it is trivial to show that the ``effective equation of state'' for $\| \bm{\Pi} \|$ (Eq.~\ref{eqn:euler}) is $\| \bm{\Pi} \| \propto \rho^{5/3}$ under isotropic compression/expansion. Note that this also imposes an additional timestep criterion, $\Delta t_{a} < C_{\rm CFL}\,h_{a} / c_{u}$ where $c_{u}^{2} = {\gamma_{\Pi}\,(\gamma_{\Pi}-1)\,\Pi^{u}_{a}/\rho_{a}}$, but this is almost always irrelevant compared to the stricter Eq.~\ref{eqn:timestep}.

This has a very direct physical interpretation. Imagine a field $\rho({\bf x},\,t)$ with a single Fourier-mode perturbation that is compressed by rapid inflow (e.g.\ two streams colliding, in \S~\ref{sec:test:nonlinear}), ignoring self-gravity. This should reach some maximum compression, where ${\bf u} \rightarrow 0$, so the energy is entirely ``stored'' in $Q$, then it will ``bounce'' back. At maximum compression, the specific energy $e = E/M \sim Q \sim \nu^{2}\,(\nabla^{2}\sqrt{\rho})/\sqrt{\rho} \sim \nu^{2}\,k^{2}$, i.e.\ if there was a large initial kinetic energy, there must be small-scale (high-$k$) modes in $Q$. Since $Q$ depends {\em only} on $k$, if the initial kinetic energy of compression (or e.g.\ gravitational force driving the compression) is sufficiently large, then eventually the required $k$ must exceed $\sim 1/\Delta x$, the numerical resolution. Regardless of the details of the numerical dissipation, {\em any} numerical method with finite resolution will fail at some point to capture the small-scale oscillations/gradients of the density field ($|\nabla \rho|/\rho$ cannot exceed $\sim 1/\Delta x$), but since the flow must still reach a point where ${\bf u}=0$, this means, naively, that energy will be lost somewhere and the ``return'' or ``bounce'' will be less energetic. The $\Pi^{u}$ term stores the energy of ``unresolved oscillations'' in the density field -- i.e.\ the un-resolved high-$k$ modes that contribute to $Q$, and therefore {\em should} contribute to the quantum pressure. In this example, when two parcels with velocities $+{\bf u}$ and $-{\bf u}$ intersect within a resolution element (so ${\bf u}\rightarrow 0$), the kinetic energy which is no longer present (and should go into sub-grid-scale oscillations of $\rho$) is correctly retained in $\Pi^{u}$, which then, correctly, ``pushes'' alongside the {\em resolved} quantum pressure from the numerically-calculated $\nabla\rho$. 

It is easy to show that if the maximum $k$ (smallest oscillation scale) is resolved, then the $\bm{\Pi}_{u}^{\ast}$ terms are completely negligible (as they should be). The field $\Pi^{u}$ therefore also serves a practical numerical purpose: by comparing $V_{a}\,\Pi^{u}_{a}$ to the total {\em resolved} energy (kinetic plus gravitational plus $Q$ computed from the resolved density gradients), one immediately obtains an estimate of convergence (specifically, the fraction of energy which has gone into un-resolved modes). 
For all the test problems in \S~\ref{sec:test:accuracy}-\ref{sec:test:nonlinear}, $k \ll 1/\Delta x$ is trivially satisfied (the mode structure is resolved); so removing the $\bm{\Pi}_{u}^{\ast}$ terms makes no visible difference whatsoever to the results shown in Fig.~\ref{fig:tests}. However, these are simple test problems with well-defined characteristic wavenumbers. For highly non-linear cosmological simulations, where external forces (gravity) are often strongly dominant over the quantum pressure, and especially when one wishes to consider larger boson masses $m$ (smaller de Broglie wavelengths), it can be important to account for the unresolved terms. Otherwise, we find that the (numerically) dissipated energy can lead to excessive growth of the central solitons at halo centers, which only converges away at extremely high (often impractical) resolution.

Of course, information is still necessarily lost when the oscillations become sub-resolution. Not only does the contribution of different $k$ modes become un-resolved, but also, in the simple approach here, the fact that we only record a {\em scalar} $ \bm{\Pi}^{\ast}_{\rm diss} $ means that we have also lost  direction information. In 3D, if we compress the field to an  unresolved peak in one direction,  the restoring force will act isotropically. In future work, we will explore whether we can improve on this by replacing $ \bm{\Pi}^{\ast}_{\rm diss} $ with an appropriate tensor that records the un-resolved components of  the compression along each direction.

\end{appendix}

\end{document}